\documentclass[12pt,superscriptaddress,showpacs]{revtex4}
\usepackage{amssymb}
\usepackage{bm}
\usepackage{graphicx}
\usepackage{psfrag}
\usepackage{color}

\def\buch{Institute for Nuclear Physics and Engineering, Bucharest, Romania}
\def\buda{KFKI Research Institute for Particle and Nuclear Physics,
  Budapest, Hungary}
\def\cler{Laboratoire de Physique Corpusculaire, IN2P3/CNRS,
 and Universit\'{e} Blaise Pascal, Clermont-Ferrand, France}
\def\darm{Gesellschaft f\"{u}r Schwerionenforschung, Darmstadt, Germany}
\def\dres{Institut f\"{u}r Strahlenphysik, Forschungszentrum Dresden-Rossendorf, Dresden, Germany}
\def\heid{Physikalisches Institut der Universit\"{a}t Heidelberg,
  Heidelberg, Germany}
\def\mosc{Institute for Theoretical and Experimental Physics, Moscow, Russia}
\def\kurc{Kurchatov Institute, Moscow, Russia}
\def\seou{Korea University, Seoul, Korea}
\def\stra{Institut de Recherches Subatomiques and
 Universit\'{e} Louis Pasteur, Strasbourg, France}
\def\wars{Institute of Experimental Physics, Warsaw University,
  Warsaw, Poland}
\def\zagr{Ru{d\llap{\raise 1.22ex\hbox
    {\vrule height 0.09ex width 0.2em}}\rlap{\raise 1.22ex\hbox
    {\vrule height 0.09ex width 0.06em}}}er
    Bo\v{s}kovi\'{c} Institute, Zagreb, Croatia}
\def\lanz{Institute of Modern Physics, Chinese Academy of Sciences,
  Lanzhou, China}

\begin{document}

\title[]{$K^0$ and $\Lambda$ production in Ni~+~Ni collisions near threshold}

\author{M.~Merschmeyer}\email{merschm@physi.uni-heidelberg.de}\affiliation{\heid} 
\author{X.~Lopez}\email{X.Lopez@gsi.de}\affiliation{\darm}
\author{N.~Bastid} \affiliation{\cler}
\author{P.~Crochet} \affiliation{\cler} 
\author{N.~Herrmann} \affiliation{\heid} 
\author{A.~Andronic} \affiliation{\darm}
\author{V.~Barret} \affiliation{\cler}
\author{Z.~Basrak} \affiliation{\zagr}
\author{M.L.~Benabderrahmane} \affiliation{\heid}
\author{R.~\v{C}aplar} \affiliation{\zagr} 
\author{E.~Cordier} \affiliation{\heid} 
\author{P.~Dupieux} \affiliation{\cler}
\author{M.~D\v{z}elalija} \affiliation{\zagr} 
\author{Z.~Fodor} \affiliation{\buda}
\author{I.~Ga\v{s}pari\'{c}} \affiliation{\zagr}
\author{Y.~Grishkin} \affiliation{\mosc} 
\author{O.N.~Hartmann} \affiliation{\darm} 
\author{K.D.~Hildenbrand} \affiliation{\darm} 
\author{B.~Hong} \affiliation{\seou}
\author{T.I.~Kang} \affiliation{\seou}
\author{J.~Kecskemeti} \affiliation{\buda} 
\author{Y.J.~Kim}\affiliation{\seou}\affiliation{\darm}
\author{M.~Kirejczyk} \affiliation{\wars} 
\author{M.~Ki\v{s}} \affiliation{\darm}\affiliation{\zagr}
\author{T.~Matulewicz} \affiliation{\wars}
\author{P.~Koczon} \affiliation{\darm} 
\author{M.~Korolija} \affiliation{\zagr} 
\author{R.~Kotte} \affiliation{\dres} 
\author{A.~Lebedev} \affiliation{\mosc} 
\author{Y.~Leifels} \affiliation{\darm}
\author{A.~Mangiarotti} \affiliation{\heid} 
\author{D.~Pelte} \affiliation{\heid}
\author{M.~Petrovici} \affiliation{\buch} 
\author{F.~Rami} \affiliation{\stra}
\author{W.~Reisdorf} \affiliation{\darm}
\author{M.S.~Ryu} \affiliation{\seou}
\author{A.~Sch\"{u}ttauf} \affiliation{\darm}
\author{Z.~Seres} \affiliation{\buda}
\author{B.~Sikora} \affiliation{\wars} 
\author{K.S.~Sim} \affiliation{\seou}
\author{V.~Simion} \affiliation{\buch}
\author{K.~Siwek-Wilczy\'{n}ska} \affiliation{\wars}
\author{V.~Smolyankin} \affiliation{\mosc} 
\author{G.~Stoicea} \affiliation{\buch} 
\author{Z.~Tyminski} \affiliation{\wars} 
\author{K.~Wi\'{s}niewski} \affiliation{\wars} 
\author{Z.G.~Xiao} \affiliation{\heid}
\author{H.S.~Xu} \affiliation{\lanz}
\author{I.~Yushmanov} \affiliation{\kurc}
\author{X.Y.~Zhang} \affiliation{\lanz}
\author{A.~Zhilin} \affiliation{\mosc}

\collaboration{FOPI Collaboration}
\noaffiliation

\date{\today}

\begin{abstract}
New results concerning the production of neutral strange particles, 
$K^0$ and $\Lambda$ in Ni~+~Ni collisions at 1.93$A$~GeV, measured
with the FOPI detector at GSI-Darmstadt, are presented. Rapidity
density distributions and Boltzmann slope parameter distributions are
measured in nearly the full phase space of the reaction. The
observables are compared to existing $K^+$ and proton data. While the
$K^0$ data agree with previously reported $K^+$ measurements, the
$\Lambda$ distributions show a different behavior relative to that of
protons. The strangeness balance and the production yield per
participating nucleon as a function of the centrality of the reaction
are discussed, for the first time at SIS energies.
\end{abstract}

\pacs{25.75.-q, 25.75.Dw}

\maketitle

\section{\label{sec:intro}Introduction}
Strangeness production and propagation in relativistic heavy-ion 
collisions are active research topics for experimental and theoretical 
nuclear physics since they are expected to provide interesting opportunities 
for studying hot and dense nuclear matter. It allows to address 
fundamental aspects of nuclear physics such as the nuclear Equation of 
State \cite{aic85,li95,stu01} and the question whether hadronic properties 
undergo modifications in such an environment \cite{ko97,cas99,fuc06}.
Moreover, it is also of crucial interest to improve our understanding
of the reaction mechanisms governing these collisions \cite{fuc97}. 
The field of heavy-ion physics is also of great importance for astrophysics, 
in particular to investigate the characteristics of the core of neutron 
stars \cite{web01}. 

The SIS energy range 1--2$A$~GeV is best suited to study the in-medium 
properties of strange particles since they are produced below or close 
to threshold. The density of the nuclear system created during the 
reaction is expected to reach up to three times normal nuclear matter 
density and its temperature is of about 90 MeV \cite{aic91,hon98}. 
Theoretical works predict that chiral symmetry can be partially
restored under those conditions, leading to changes of hadron
properties \cite{ko96,ko97} affecting both, production and
propagation. Indications for in-medium modifications of charged kaons
have been already observed experimentally with data from FOPI
\cite{cro00,wis00} and KaoS \cite{kaos}.

Neutral strange particles such as $K^0$ and $\Lambda$ are of great
interest because of their associated production and the fact that they
are not influenced by the Coulomb interaction. The study of their
properties may probe the in-medium potential, which is expected to be
weakly repulsive for $K^+$ ($K^0$) \cite{li97} and attractive for
$\Lambda$ \cite{wan99}. It is worth to point out that such opportunity
cannot be offered by studies of hypernuclei \cite{mil88} which allow
to investigate the $\Lambda$-nucleon potential at normal nuclear
matter density.

Detailed studies of the properties of these neutral strange particles
have been performed at the LBL BEVALAC and the BNL Alternating Gradient
Synchrotron (AGS) \cite{jus98,chu00,chu01,alb02,chu03}.
In this paper new results on the production of $K^0$ and
$\Lambda$ in Ni~+~Ni collisions at 1.93$A$~GeV, near threshold (which
is 1.58~GeV for free $NN$ collisions), are presented. Differential
production yields and inverse slope parameters from close to the full
phase space are discussed. In particular the high statistics collected
during the experiment allows for the first time to investigate the
centrality dependence of the $K^0$ and $\Lambda$ production yields at
SIS energies.

The paper is structured as follows. Section \ref{sec:exprm} consists
out of a short description of the apparatus. Section \ref{sec:cent} is
devoted to the event characterization, namely centrality and
azimuthal orientation of the reaction plane. The reconstruction method
is detailed in section \ref{sec:recns}. Section \ref{sec:reslt}
addresses the experimental results, yields and effective (Boltzmann)
slope parameters of $K^0$ and $\Lambda$, which are also compared to
existing data of $K^+$ and proton. Subsequently the strangeness
balance and the centrality dependence of $K^0$ and $\Lambda$
production are discussed. Finally, a summary and an outlook are given
in section \ref{sec:concl}.

\section{\label{sec:exprm}Experiment}
The experiment was performed at the GSI Schwerionen Synchrotron (SIS) in
Darmstadt (Germany) by bombarding a $^{58}_{28}$Ni beam of 1.93$A$~GeV
on an enriched target ($>$95\%) of $^{58}_{28}$Ni. The target
thickness was about 360~mg/cm$^2$ corresponding to an interaction
probability of 1.5\% and the average beam intensity was
4\ --\ 5x10$^5$~ions/s.

The FOPI setup is an azimuthally symmetric apparatus made of several 
sub-detectors which provide charge and mass determination over nearly the
full 4$\pi$ solid angle. The central part ($ 23^\circ < \theta_{lab} <
113^\circ$) is placed in a super-conducting solenoid and comprises a
drift chamber (CDC) surrounded by a barrel of plastic scintillators
for polar angles between $ 32^\circ$ and $ 72^\circ$. The mass of
particles measured in the CDC is determined using magnetic rigidity
and energy loss. The forward part is composed of a wall of plastic
scintillators ($1^\circ<\theta_{lab}<25^\circ$) and another drift
chamber (Helitron) mounted inside the super-conducting solenoid. The
plastic wall is divided in two parts: the inner plastic wall and the
outer plastic wall (PLAWA) which cover polar angles
$1^\circ<\theta_{lab}<6.6^\circ$ and $6^\circ<\theta_{lab}<25^\circ$, 
respectively. The plastic wall provides charge identification of the
reaction products, combining time of flight and specific energy loss
informations. For the present analysis, the PLAWA and the CDC were
used. More details on the configuration and performances of the
different components of the FOPI apparatus can be found in
\cite{gob}.

During the experiment, more than 110x10$^6$ events were recorded
under ``central'' and ``medium-central'' triggers, corresponding to
different thresholds on the multiplicity of charged fragments measured
in the PLAWA. Most of the events (100x10$^6$) were registered
under the ``central'' trigger which accepts events covering up to
about 20\% of the total reaction cross section.

In previous experiments \cite{bes97,hon98,cro00,wis00} the target was
located at a different position +40~cm in beam direction which led
to a better acceptance around target rapidity. With our present target
setup, the whole backward hemisphere is covered.

\section{\label{sec:cent}Event Characterization} 
\subsection{Centrality Selection}
The events are sorted according to their degree of centrality by
imposing cut conditions on the total charged particle multiplicity
measured in both, PLAWA and CDC (i.e.~MUL).
\begin{table}[!hbt]
 \begin{tabular}{c @{\quad} c @{\quad} c @{\quad} c @{\quad} c @{\quad} c}
 \hline
 \hline
        & $\sigma_{geo,min}$ & $\sigma_{geo,max}$ &
 $\Delta\sigma_{geo}$ &  $\left<b_{geo}\right>$ & $A_{part}$ \\
  Class & (mb) & (mb) & (mb) & (fm) & \\
 \hline
  MUL1 &   0 &   98 &  98 & 1.7 $\pm$ 0.4 & 94 $\pm$ 6 \\
  MUL2 &  98 &  389 & 291 & 3.6 $\pm$ 0.4 & 63 $\pm$ 6 \\
  MUL3 & 398 &  679 & 290 & 4.7 $\pm$ 0.3 & 43 $\pm$ 5 \\ 
  MUL4 & 679 & 1005 & 326 & 5.5 $\pm$ 0.3 & 32 $\pm$ 5 \\
 \hline
  Central & 0 & 350 & 350 & 3.1 $\pm$ 0.4 & 71 $\pm$ 6 \\
 \hline
 \hline
 \end{tabular}
 \caption{\label{tab-1} Cross section, mean geometrical impact
 parameter, and number of participants for Ni~+~Ni collisions at
 1.93$A$~GeV for five different centrality selections (see text for
 details).}
\end{table}  

Four event classes are used for the investigation of the centrality
dependence of the $K^0$ and $\Lambda$ production yields
(Tab.~\ref{tab-1}, labeled ``MUL1'' to ``MUL4''). The characteristics
of the central event class used to study the $K^0$ and $\Lambda$
differential distributions are provided in the last row of
Tab.~\ref{tab-1} (labeled ``Central'').
The corresponding ranges (from $\sigma_{geo,min}$ to
$\sigma_{geo,max}$) and the portions of the geometrical cross section
of the centrality classes ($\Delta\sigma_{geo}$) are listed. The mean
geometrical impact parameter $\left<b_{geo}\right>$ and its
r.m.s.\ deviations are determined assuming a sharp-cut-off
approximation \cite{cav90}. We calculate the average number of
nucleons in the fireball ($A_{part}$) using the geometrical model
described in \cite{gos77}.

\subsection{Reaction Plane} 
When applying the event-mixing technique for the combinatorial
background determination one needs to rotate all events into the
reaction plane (see section IV). This plane is estimated on an
event-by-event basis according to the standard transverse momentum
procedure devised in \cite{dan85} which allows to construct the
event-plane vector
\begin{equation}
\label{eqt1}
\vec{Q}=\sum_\nu \omega_\nu \vec{u}_\nu \textrm{ .} 
\end{equation}
The sum runs over all charged particles measured in the PLAWA and the
CDC. The unit vector $u_\nu$ is parallel to the particle transverse
momentum (i.e. $u_\nu~=~(\cos\varphi_\nu,\sin\varphi_\nu)$, where
$\varphi_\nu$ is the particle azimuth), and $\omega_\nu$ is a weight
to improve the resolution, depending on the center-of-mass (c.m.)
rapidity $y_{c.m.}= y_{lab} - 0.5\cdot y_p$ (the subscript $p$ refers to
the projectile): $\omega_\nu=-1$ for $y_{c.m.}<-0.2$, $\omega_\nu=+1$
for $y_{c.m.}>+0.2$ and $\omega_\nu=0$ otherwise. The azimuth of $Q$ is
an estimate of the azimuth of the reaction plane. Typically, the
resolution on the reaction plane determination estimated as in
\cite{dan85} from randomly chosen sub-events, is about $30^\circ$ for
central events ($\sigma_{geo}~\approx$~350~mb, $b_{geo}~<$~4.0~fm).

\section{\label{sec:recns}Particle Reconstruction}
Long-lived neutral strange particles like the $K^0$ or the $\Lambda$
are identified via their charged decay particles in the CDC. $K^0_L$
and $K^0_S$ are produced in equal amounts due to the nature of the
weak interaction (neglecting $CP$ violation).
Because of its lifetime and decay channels the $K^0_L$ cannot be
measured with the FOPI apparatus, whereas the $K^0_S$ can be
reconstructed from its decay into $\pi^-$ and $\pi^+$ (branching
ratio~=~68.6\%). Similarly, the reconstruction of the $\Lambda$ is
done from the final state containing $\pi^-$ and $p$ (branching
ratio~=~63.9\%). A possible contribution to the observed $\Lambda$
comes from the $\Sigma^0$ which decays exclusively into a $\Lambda$
and a photon. Since the photon cannot be measured with the FOPI
detector, $\Lambda$ and $\Sigma^0$ cannot be disentangled and are
referred to as $\Lambda$ in the following.

The long lifetimes of the weakly decaying $K^0_S$ ($c\tau~=~2.68$~cm)
and the $\Lambda$ ($c\tau~=~7.89$~cm) causes a sizable fraction of
these particles decaying somewhat away from the primary vertex. The
precision of the track reconstruction in the CDC is sufficient to
resolve these secondary vertices. Therefore the signal-to-background
ratio is improved significantly by rejecting particles coming from the
primary vertex.

\subsection{\label{ssec:meth}Reconstruction Method}
The search for $K^0_S$ and $\Lambda$ starts with the selection of
possible final state particles. Three cuts for each track in the CDC
are applied. The first one rejects particles with transverse momenta
lower than $80-100$~MeV/c which are spiraling in the CDC.
The particle species is selected using the correlation of the particle
mass, which is determined from the measured specific energy loss, and
the total momentum.
The distance of closest approach between track and primary event
vertex in the transverse plane is the third quantity. It is used to
enhance the fraction of particles coming from secondary vertices in
the data sample.

\begin{figure}[!ht]
 \includegraphics[width=0.8\columnwidth]{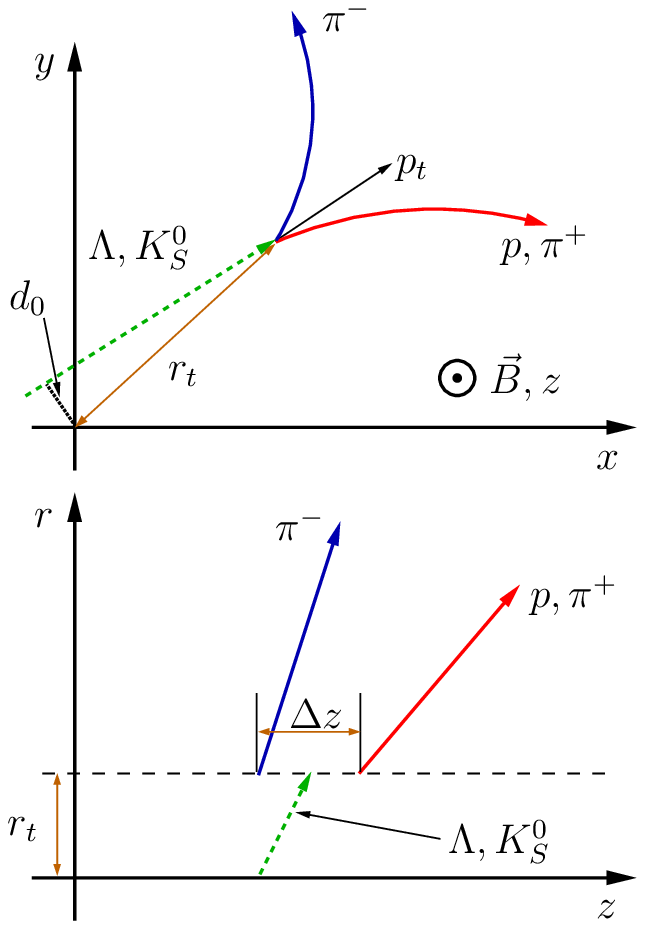}
 \caption{\label{fig:recs} (Color online) Schematic view of the $K^0_S$ and 
 $\Lambda$ reconstruction in the $x-y$ plane (top) and in the 
 $r-z$ plane (bottom). 
 The cut quantities used for the selection of the candidates are 
 illustrated in the plots.}
\end{figure}

For a jet-type drift chamber like the CDC it is advantageous to pursue
a reconstruction strategy which first focuses on the transverse plane
($x-y$) in order to find intersection points of possible decay
particle tracks as illustrated in the upper plot of
Fig.~\ref{fig:recs}. Once such a point is found, further cut
quantities are determined.
The transverse distance to the primary vertex $r_t$ and the closest
approach of the back-extrapolated mother particle flight path 
$d_0$ are used to suppress the background.
Then the quality of the signal can be improved by taking into account
the longitudinal distance $\Delta z$ of the decay particle tracks as
shown in the lower plot of Fig.~\ref{fig:recs}.
Afterwards, the invariant mass of the particle pair is calculated from
its four-momenta at the intersection point.

\subsection{\label{ssec:invm}Invariant Mass Spectra and Phase-Space Distributions}
\begin{figure}[!ht]
 \includegraphics[width=\columnwidth]{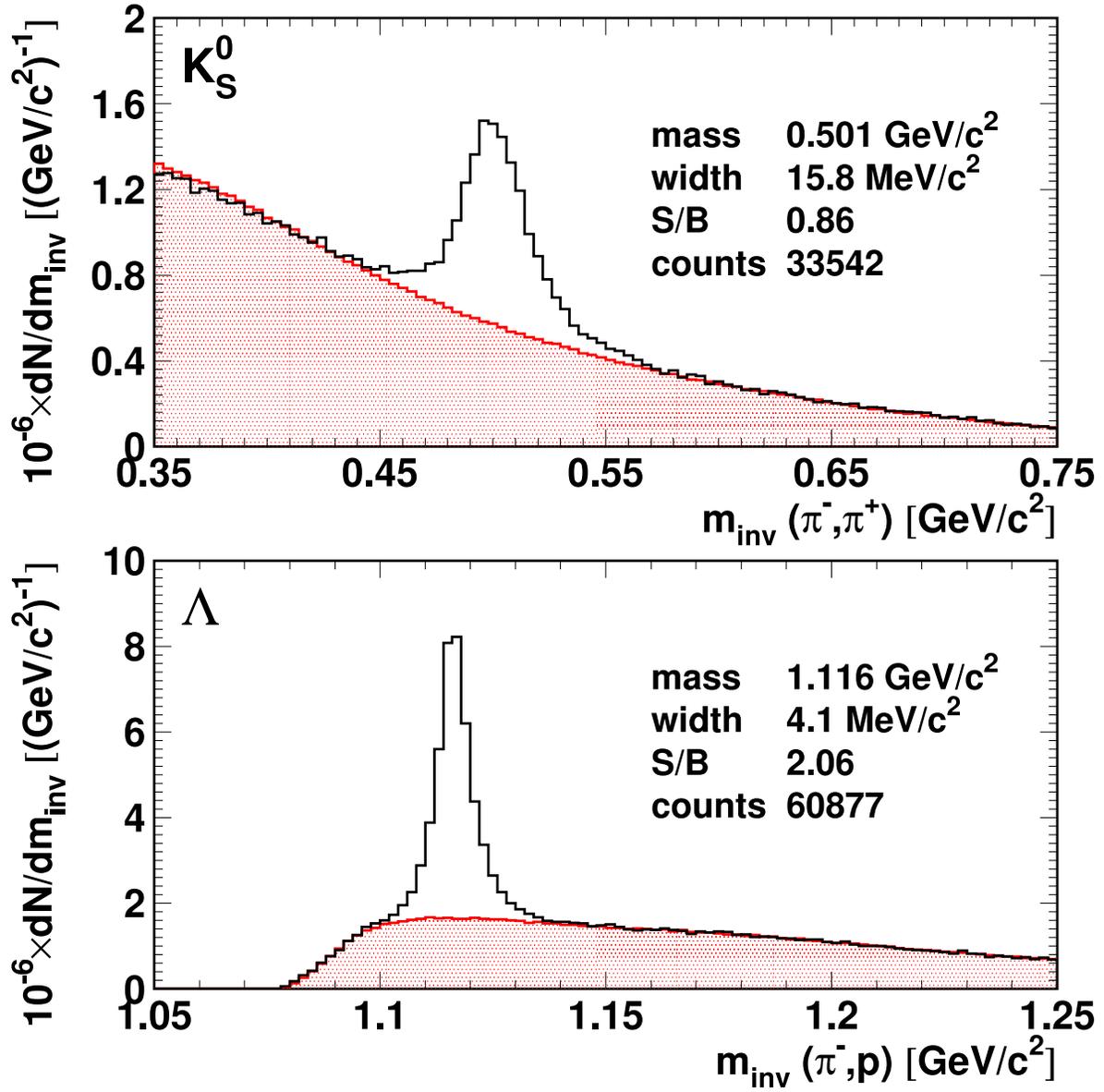}
 \caption{\label{fig:minv} (Color online) Invariant mass spectra of $\pi^-\pi^+$ (top) and 
  $\pi^- p$ (bottom) pairs. The solid lines denote the
  combinatorics, the shaded areas represent the scaled
  mixed-event background.}
\end{figure}

The invariant mass spectra of $\pi^-\pi^+$ and $\pi^- p$ pairs for a
centrality selection of the most central 350~mb of the geometrical
cross section are shown in Fig.~\ref{fig:minv}. A total of about
47x10$^6$ events were analyzed.
The combinatorial background was modeled with the event-mixing method
\cite{ber77}. For that purpose the two decay particle candidates are
taken from two different events with the same centrality.
In addition, the two events are projected into the same reaction plane. 
Afterwards, a normalization is applied to the mixed-event background
in order to fit the combinatorics in the invariant mass range
$0.56-0.80$~GeV/c$^2$ and $1.15-1.25$~GeV/c$^2$ for $K^0_S$ and
$\Lambda$, respectively. The shape of the resulting mixed-event
background fits nicely to that of the combinatorial background and is
indicated by the shaded area in Fig.~\ref{fig:minv}.
After background subtraction, the remaining peaks in the mass spectra
are fitted with Gaussians. Within an interval of $\pm2\sigma$ around
the center of the peak, about 33000 $K^0_S$ and 61000 $\Lambda$
candidates are found for the applied set of cuts in the analysis.
The signal-to-background ratios are of 0.9 ($K^0_S$) and 2.1 ($\Lambda$). 
The (Gaussian) widths of 15.8~MeV/$c^2$ ($K^0_S$) and 4.1~MeV/$c^2$
($\Lambda$) are resulting from the detector resolution.

\begin{figure}[!ht]
 \includegraphics[width=\columnwidth]{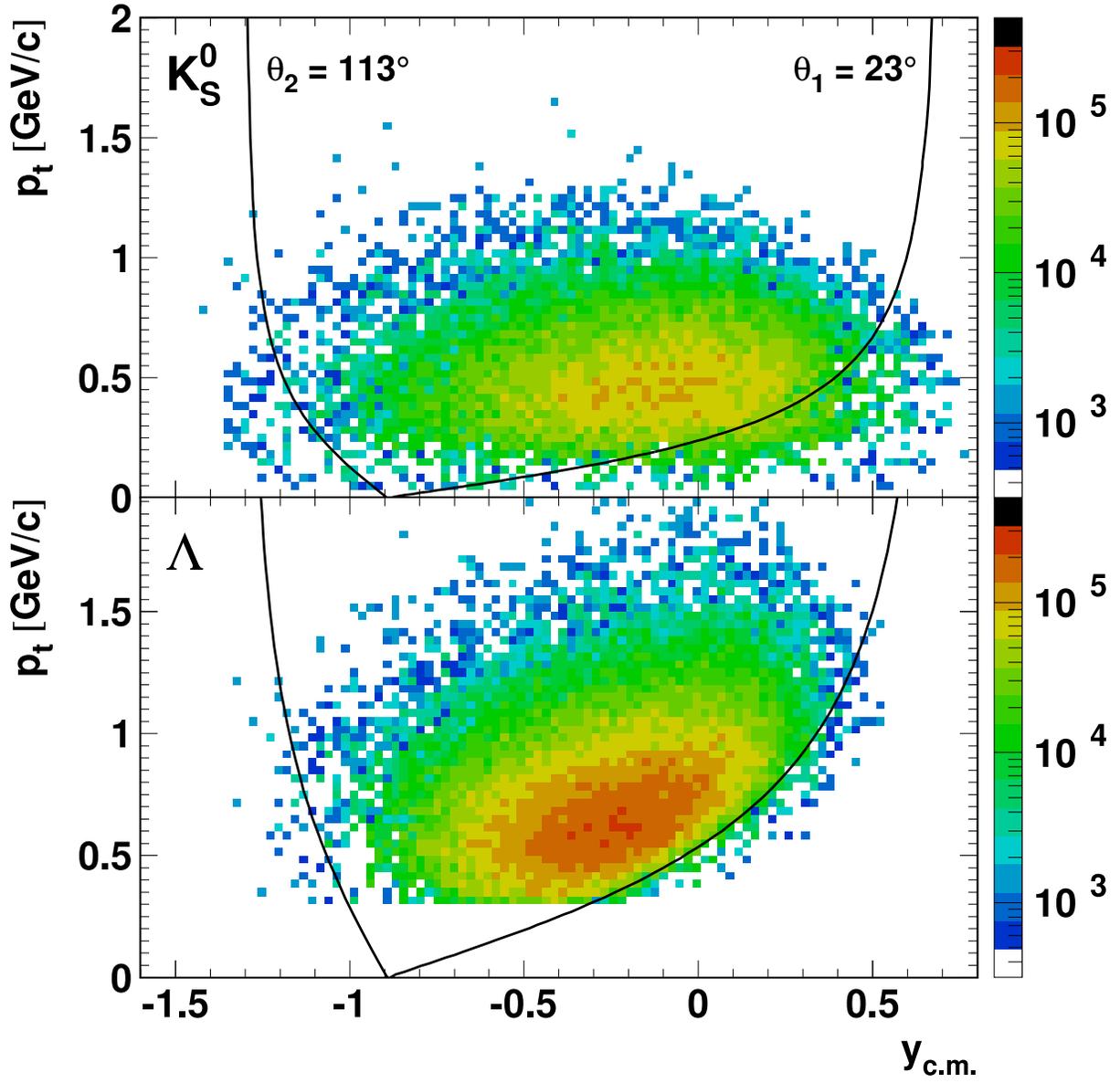}
 \caption{\label{fig:phsp} (Color online) Phase space distributions of the reconstructed
  $K^0_S$ (top) and $\Lambda$ (bottom), in the plane $p_t$ {\it vs.}
  $y_{c.m.}$. The solid lines indicate the polar angular coverage of the CDC.}
\end{figure}

The background-subtracted phase-space distributions of those particle
candidates are presented in Fig.~\ref{fig:phsp} where the transverse
momentum $p_t$ is plotted versus the center-of-mass rapidity
$y_{c.m.}$. The solid lines denote the limits of the CDC's polar angular
acceptance of 23$^\circ$ and 113$^\circ$ in the forward and backward
directions, respectively.
Both distributions illustrate the nearly complete coverage of the
backward hemisphere. By using the symmetry of the colliding system,
most of the phase space becomes accessible.
In addition, the $K^0_S$ distribution shows a fraction of candidates
even coming from outside the CDC acceptance. This is possible because
both decay particles can still reach the CDC acceptance due to the
high excess energy in the decay process.

\subsection{\label{ssec:effc}Acceptance and Efficiency Correction}
In order to extract global quantities like the particle yields from
the data, the losses due to acceptance and efficiency have to be
corrected. For a complex apparatus like the FOPI detector these
corrections have to be determined by an extensive GEANT simulation
modeling all detector components.
The particle distributions of heavy-ion collisions are generated by
the IQMD model \cite{aic91,har98}. The neutral strange particles
($K^0_S,~\Lambda$) were added separately using a momentum
distribution given by the Siemens-Rasmussen formula \cite{sie79} which
describes a system expanding with a temperature $T$ and a radial
expansion velocity $\beta$. The choice of the parameters ($T=90~MeV$,
$\beta=0.3$) is suggested by previous results obtained for $\pi^-$,
protons and deuterons \cite{hon98}. $K^0_S$ and $\Lambda$ particles
were then embedded into the IQMD output with one $K^0_S$ and one
$\Lambda$ per event. A full simulation of the detector including
resolutions in energy deposition and spatial position, front-end
electronics processing and hit tracking, is performed and the
resulting output is subject to the same reconstruction procedure as
the experimental data. Spectra of all relevant cut quantities were
compared between simulated and experimental data
\cite{mm04,lop04}. Since no significant differences were found, one
can conclude that the detector is properly described by the
simulation.

Finally, the reconstruction efficiency is determined by computing the
ratio of the number of reconstructed particles in the simulation to
the number of particles embedded into the input events. This is done
in bins of rapidity $y_{c.m.}$ and transverse mass $m_t-m_0$, where
$m_t=\sqrt{m^2_0+p^2_t}$ and $m_0$ is the rest mass of the
particle. Thus, the correction method is independent of the choice of
the source temperature for the embedded strange particles.

\begin{figure}[!ht]
 \includegraphics[width=\columnwidth]{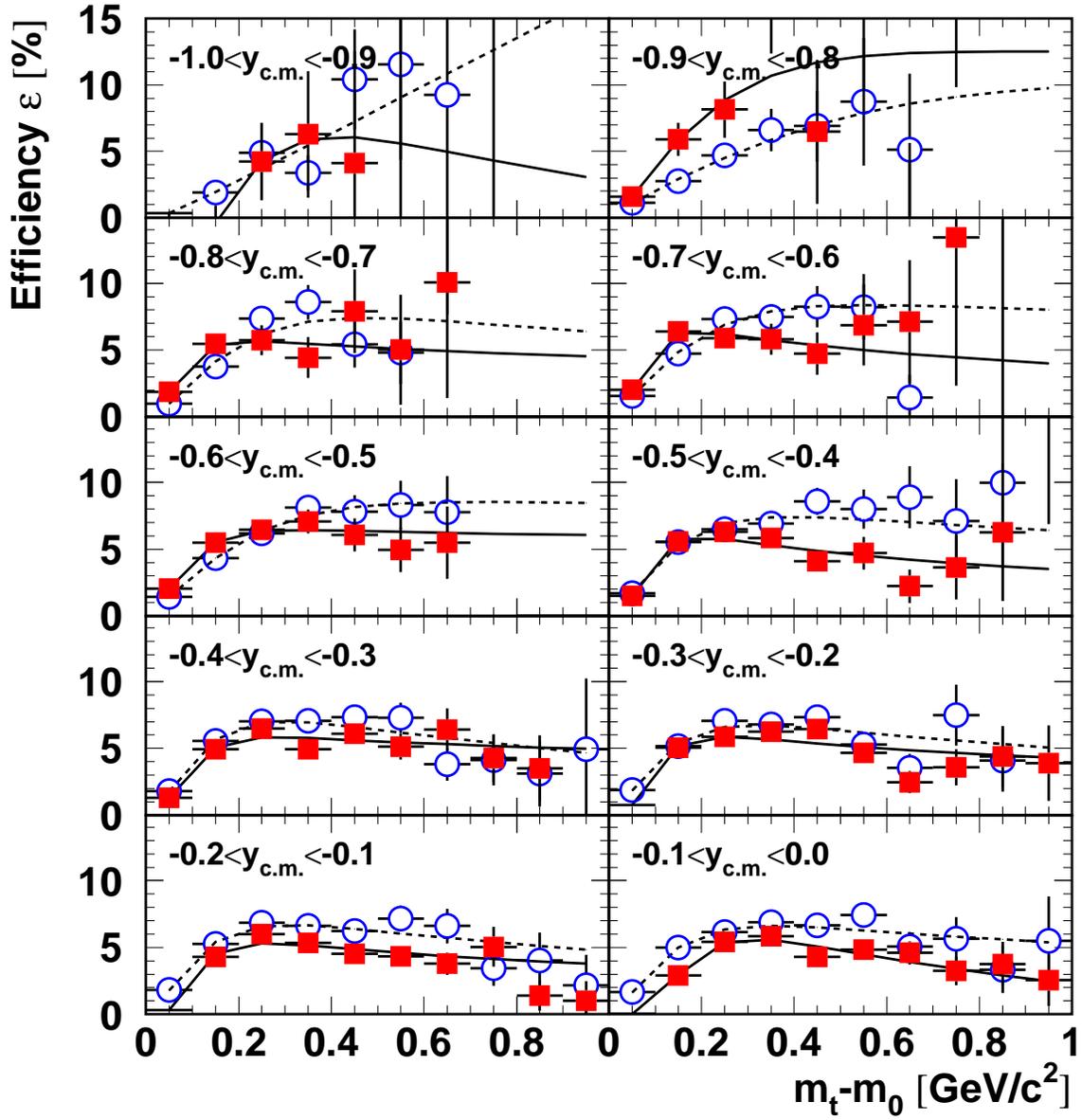}
 \caption{\label{fig:effc} (Color online) Reconstruction efficiencies for the $K^0_S$
   (open circles) and the $\Lambda$ (full squares) as a function of
   the transverse mass $m_t-m_0$ given for different bins of rapidity
   $y_{c.m.}$. The dashed and solid lines denote fits to the
   efficiencies of $K^0_S$ and $\Lambda$, respectively.}
\end{figure}

The reconstruction efficiencies for $K^0_S$ (open circles) and
$\Lambda$ (full squares) are plotted in Fig.~\ref{fig:effc}. The
numbers are not corrected for the branching ratio of the respective
decay channels. Due to the fact that a minimum transverse momentum of
90--100~MeV/c is requested for the decay particles, the efficiency is
low for small values of $m_t-m_0$ ($<100$~MeV/c$^2$). At higher
transverse masses the efficiencies are approximately independent of
$y_{c.m.}$ and $m_t-m_0$ with maximum efficiencies of about 7\% for the
$K^0_S$ and 5\% for the $\Lambda$, respectively. In order to apply a
smooth efficiency correction to the data, which is less sensitive to
fluctuations in the high-$m_t$ range, the following function was
fitted to the reconstruction efficiencies:
\begin{equation}
 \label{eqt1b}
  f=a\cdot\tanh(b\cdot (m_t-m_0))+c\cdot(m_t-m_0)^d \textrm{ ,}
\end{equation}
where $a$, $b$ and $c$ are free fit parameters and $d$ has a value of
0.45 and 0.2 for $K^0$ and $\Lambda$, respectively. This function
was then scaled by the branching ratio of the decay channel and
applied to the measured transverse mass spectra.

\section{\label{sec:reslt}Results}
In this section, the transverse mass spectra of $K^0$ and $\Lambda$ 
after background subtraction and efficiency correction, are fitted
with a Boltzmann-type distribution in order to extract the rapidity
dependencies of the Boltzmann temperature, the rapidity density
distributions and ultimately, the effective temperature and the total
production yield in central (350~mb) Ni~+~Ni reactions at 1.93A~GeV.
The distributions obtained for $K^0$ and $\Lambda$ are compared to
existing data on $K^+$ and protons, respectively. The large statistics
of reconstructed $K^0_S$ and $\Lambda$ allows for the first time to
discuss the total production yield as a function of the centrality of
the reaction, as it has been done at AGS energies
\cite{jus98,chu00,chu01,chu03}.
 
In the following, the results include only statistical errors.
Note that the differential yield distribution of the $K^0_S$ was
scaled by a factor two in order to account for the unmeasured
contribution from the $K^0_L$ and is therefore referred to as $K^0$.

\subsection{\label{ssec:islpt}Transverse Mass Spectra}
\begin{figure}[!ht]
 \includegraphics[width=\columnwidth]{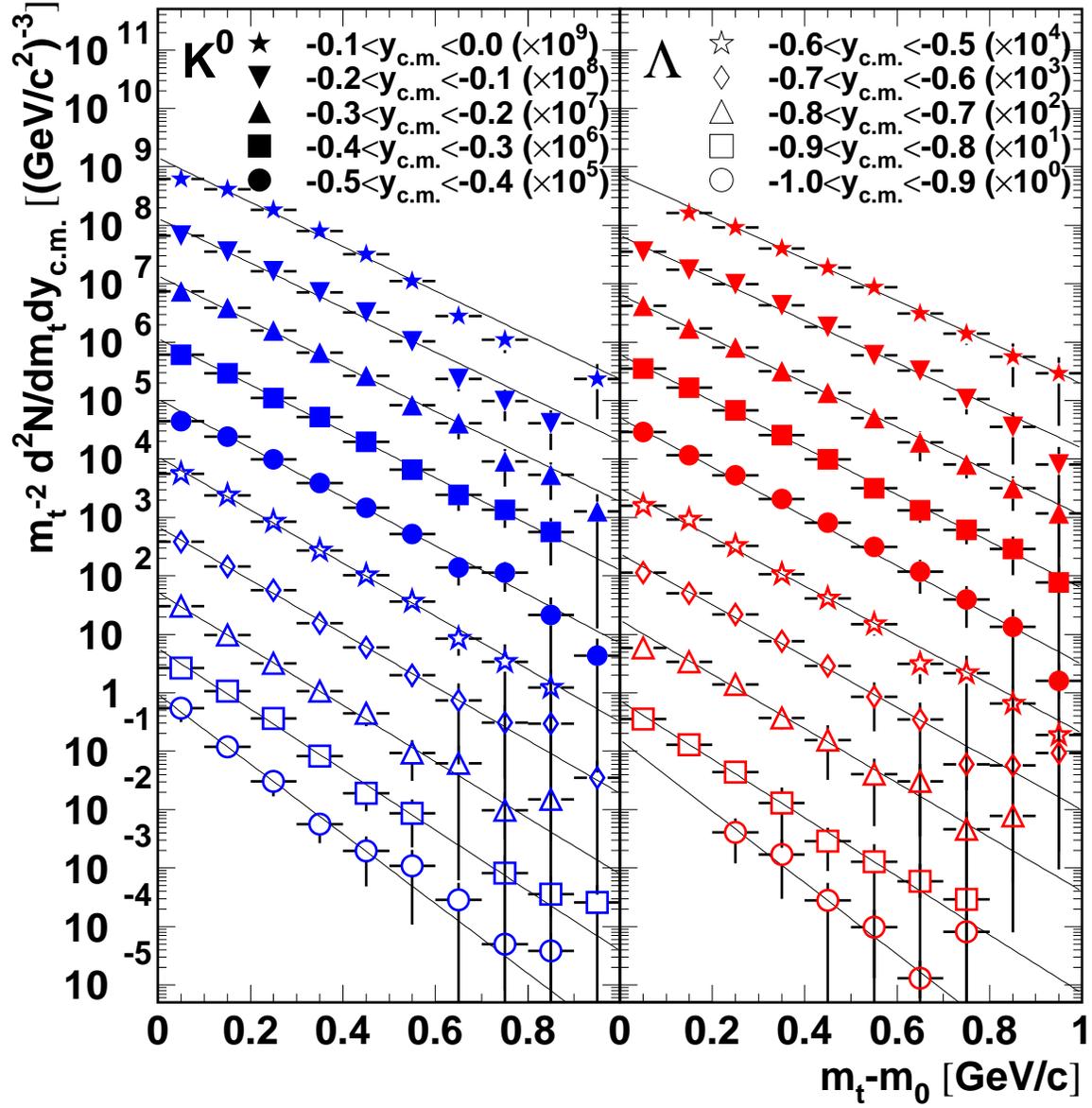}
 \caption{\label{fig:mtsp} (Color online) Transverse mass spectra of reconstructed
 $K^0$ (left) and $\Lambda$ (right). The spectra are plotted for ten
 rapidity bins ranging from $y_{c.m.}=$~-1.0 (lowermost spectrum, scaled
 by 10$^0$) to $y_{c.m.}=$~0.0 (uppermost spectrum, scaled by 10$^9$).
 The data are corrected for the branching ratios of the respective
 decay channels and the unmeasured $K^0_L$ contribution.}
\end{figure}

Figure~\ref{fig:mtsp} shows the transverse mass spectra $m_t - m_0$ of 
$K^0_S$ and $\Lambda$ particles, after background subtraction and 
correction for efficiency, branching ratios and the unmeasured $K^0_L$
contribution, for several rapidity bins indicated in the figure.
The rapidity range $-1.0~<~y_{c.m.}~<~0.0$ is covered, hence the full
phase space is accessible by exploiting the symmetry of the Ni~+~Ni
system. In order to compile all spectra in one plot they are
multiplied by $10^n$ starting from the lowermost spectrum ($n = 0$) to
the uppermost spectrum ($n = 9$).

If particles are emitted from a thermal Boltzmann-like source, their 
behavior is then described by the following function:
\begin{equation}
 \label{eqt2}
 \frac{1}{m^2_t}\frac{d^2N}{d(m_t-m_0)dy_{c.m.}} = A \cdot 
 \exp \frac{-(m_t-m_0)}{T_B}  
\end{equation}
within a narrow window of rapidity $dy_{c.m.}$. 
Both, the inverse slope parameter ($T_B$) and the integration constant ($A$) 
depend on the rapidity. This allows to extract the rapidity
distributions of the inverse slope parameter $T_B$ and the rapidity
density distributions $dN/dy_{c.m.}$ by integrating the fitted function
from $m_t-m_0$ = 0 to $\infty$. In the logarithmic representation, our
measured spectra exhibit a linear decrease with increasing $m_t-m_0$
and are described reasonably well by Eq.~\ref{eqt2}.

\subsection{\label{ssec:islp}Inverse Slope Parameters}
The inverse slope parameter $T_B$ yields information on the particle
temperature at freeze-out for a thermalized system. The Boltzmann
temperature $T_B$ has a simple dependence on the rapidity for an
isotropically emitting source:
\begin{equation}
 \label{eqt3} 
 T_B = \frac{T_{eff}}{\cosh(y_{c.m.})}, 
\end{equation}
where $T_{eff}=T_B(y_{c.m.}=0)$ is the effective temperature, i.e.~the
slope parameter at mid-rapidity.

\begin{figure}
 \includegraphics[width=\columnwidth]{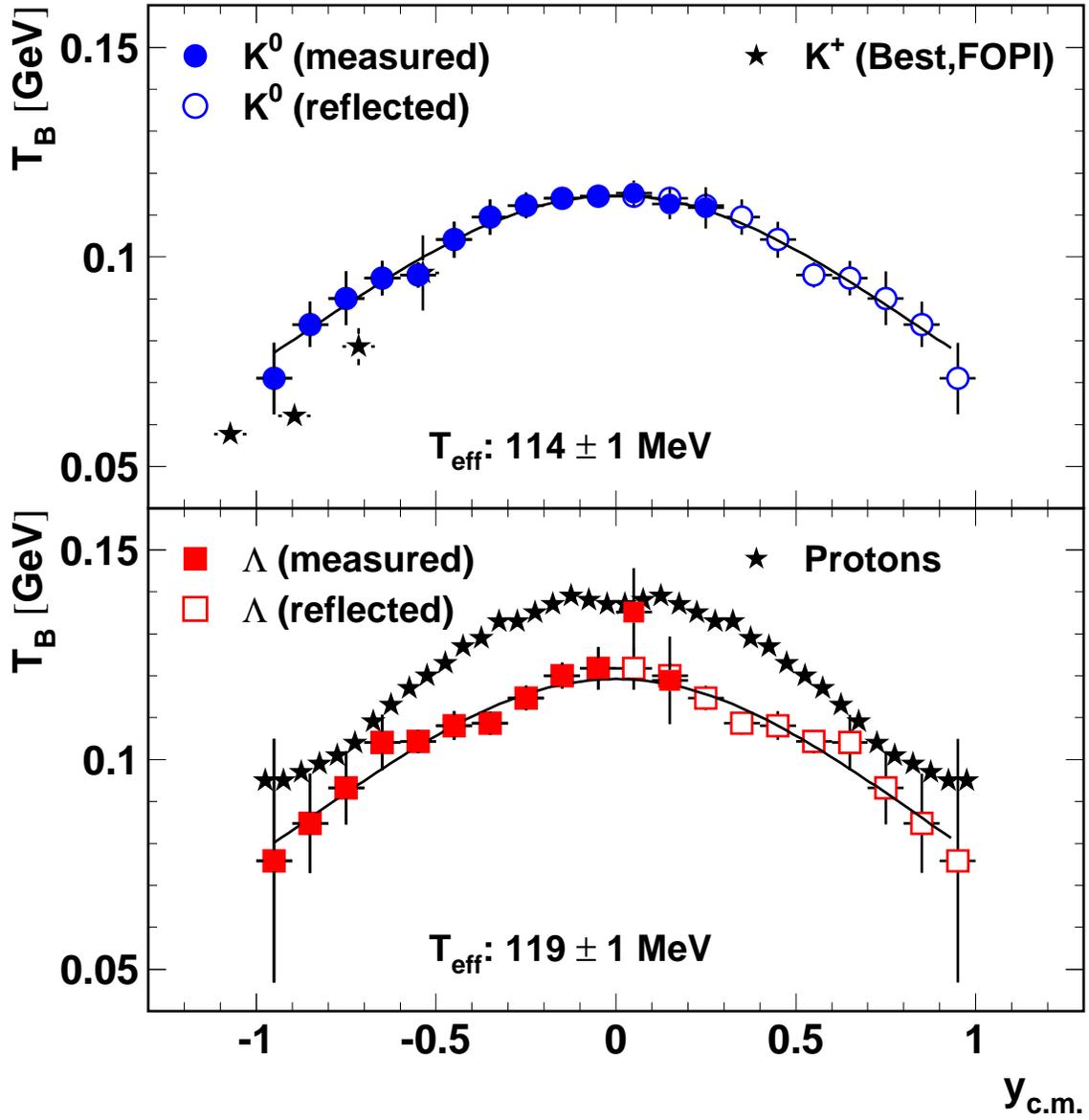}
 \caption{\label{fig:islp} (Color online)
 Rapidity dependency of the inverse slope parameter for $K^0$
 (full circles, upper plot) and $\Lambda$ (full squares, lower plot).
 The open symbols are data points reflected with respect to
 mid-rapidity. The star symbols in the upper plot denote previously
 measured $K^+$ \cite{bes97} and the star symbols in the lower plot
 denote protons from our experiment. The lines represent a fit
 assuming an isotropic thermalized source (see text for details).}
\end{figure}

The rapidity dependencies of $T_B$ are depicted in Fig.~\ref{fig:islp}
for various particle species. The full circles and squares indicate
measured data points for $K^0$ and $\Lambda$ while open symbols
denote points reflected with respect to mid-rapidity.
The star symbols correspond to previously measured distributions of 
$K^+$ from \cite{bes97} and to protons from this experiment. 
For the $K^0$ an effective temperature of 114~$\pm$~1 MeV is
extracted from the fit using Eq.~\ref{eqt3} which agrees very well
with the $K^+$ results from \cite{for07}.
The $T_B$ distribution of $K^0$ is found to be slightly higher
than the one measured for $K^+$, within the overlap region.
Note that $K^+$ were measured directly, i.e.\ via time-of-flight and
momentum, and that in previous FOPI experiments the target was in its
nominal position which explains the different rapidity range for
$K^+$.
Comparing the combined KaoS and FOPI results for the $K^+$ to the
$K^0$ suggests that they have very similar freeze-out
conditions.

The lower part of the Fig.~\ref{fig:islp} displays the corresponding
results for the $\Lambda$ with an the effective temperature of
119~$\pm$~1~MeV. The $\Lambda$ have a somewhat lower freeze-out
temperature ($\Delta T~\approx$~20~MeV) at mid-rapidity as compared to
protons. This difference in $T_B$ is present throughout most of the
phase space, only around target and projectile rapidities both
measurements agree within errors.

Similar results were reported by the EOS experiment \cite{jus98} for
Ni+Cu collisions at energies of about 2$A$~GeV. Effective temperatures
of 106~$\pm$~5~MeV ($\Lambda$) and 142~$\pm$~1~MeV (protons) have been
derived from an exponential fit to the measured transverse mass spectra
within the central rapidity range $\left|y/y_{beam}\right|\le$~0.25.

\subsection{\label{ssec:ylds}Production Yields}
The rapidity density distribution $dN/dy_{c.m.}$ is determined
separately for each bin in rapidity by an analytic integration of
Eq.~\ref{eqt2} from $m_t=m_0$ to $m_t\rightarrow\infty$, using $A$ and
$T_B$ parameters from the fit. One obtains \cite{sch93}: 
\begin{equation}
 \label{eqt6} 
 \frac{dN}{dy_{c.m.}}=A \times T^3_B \times
 \left( \frac{m_0^2}{T^2_B}+2 \frac{m_0}{T_B}+2\right),   
\end{equation}
where $m_0$ is the particle's rest mass. Using this method,
contributions to the integral from outside the acceptance are
included.

\begin{figure}[!hbt]
 \includegraphics[width=\columnwidth]{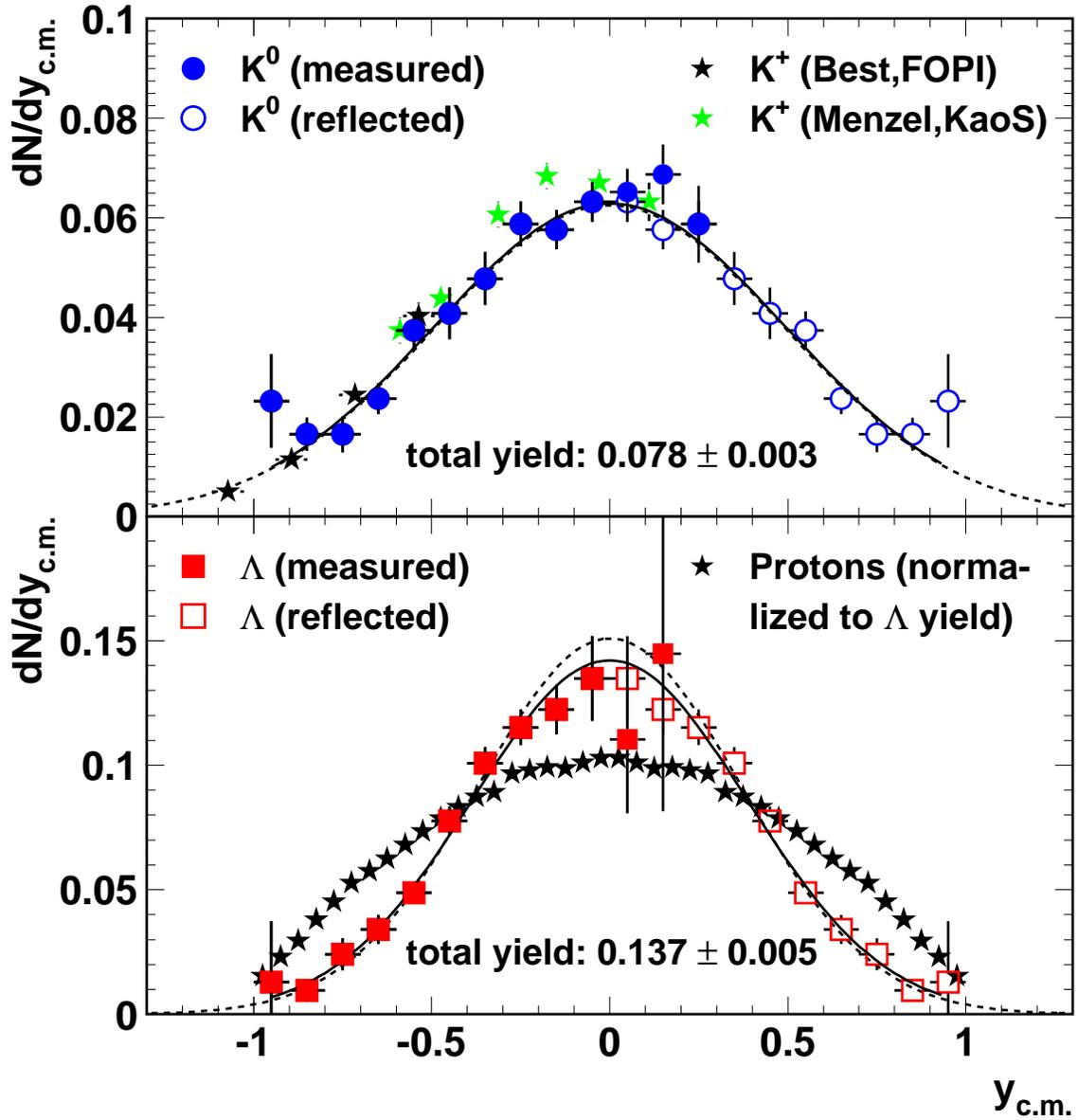}
 \caption{\label{fig:dndy} (Color online)
 Rapidity density distributions for $K^0$ (full circles, upper plot)
 and $\Lambda$ (full squares, lower plot). The open symbols are data
 points reflected with respect to mid-rapidity.
 The (black and green) star symbols correspond to previously measured
 $K^+$ \cite{bes97,men00} (upper plot) and to protons from this
 experiment (lower plot). The lines represent a Gaussian fit from
 which the total production yield is extracted (see text for
 details).}
\end{figure}

The resulting $dN/dy_{c.m.}$ distributions are presented in
Fig.~\ref{fig:dndy}. The meaning of the different symbols is the same
as in Fig.~\ref{fig:islp}.
The distributions for $K^0$ and $\Lambda$ particles are corrected
for the corresponding branching ratios.
The upper plot shows a comparison between $K^0$ and $K^+$ distributions 
\cite{bes97,men00}, which are found to be very similar. The total $K^0$ 
production yield per central (350~mb) collision of 0.078$\pm$0.003 and
the corresponding width of the $dN/dy_{c.m.}$ distribution of
0.493$\pm$0.016 are obtained from a Gaussian fit to the data. The
combined $K^+$ distribution has a width of 0.433$\pm$0.011 and a total 
production yield of 0.075$\pm$0.002.

The $dN/dy_{c.m.}$ distribution of $\Lambda$ particles is displayed in the 
lower panel of Fig.~\ref{fig:dndy} and is compared to the one of
protons (down-scaled in order to have the same integral as the
$\Lambda$ data). One observes a pronounced longitudinal spread of the
proton distribution which could be an indication for a large degree of 
transparency in the Ni~+~Ni system, which has already been found for
other systems \cite{hon02,rei04}.
The $\Lambda$ rapidity distribution is much more compact as expected
for particles produced in the fireball.
Such trend is also observed at AGS energies \cite{alb02}. The total
$\Lambda$ production yield per central (350~mb) collision is
0.137$\pm$0.005 and the width of the distribution is 0.386$\pm$0.009.

The dashed curves in both plots denote a Siemens-Rasmussen
distribution~\cite{sie79} (normalized to the respective total yield)
for a temperature of $T~=$~90 MeV and radial expansion velocity
$\beta~=$~0.3 as found in \cite{hon98}. While the kaons fully agree
with the Siemens-Rasmussen distribution, a slight difference is
visible for the $\Lambda$ in the mid-rapidity region. This may
indicate a non-isotropic emission of the $\Lambda$ in the
center-of-mass system.

\subsection{\label{ssec:syse}Evaluation of the Systematic Errors}
In order to estimate the systematic error on the width of 
the rapidity density distributions, on the effective temperatures 
and on total production yields we apply different sets of cuts by varying the 
selection criteria of the relevant quantities discussed in section
\ref{sec:recns}. These selection criteria are determined so that it is
possible to accept variations of counts in the invariant mass spectra
of about $\pm$50\%. While the effective temperatures vary within
6--8\% for both particles, the widths of the rapidity density
distributions  differ by about 6\% ($K^0$) and 12\%
($\Lambda$). The total production yields change by about 10\%
($K^0$) and 18\% ($\Lambda$). The systematic errors are summarized
in Tab.~\ref{tab-3}.

\begin{table}[!h]
 \begin{tabular}{r @{\qquad} c @{\qquad} c}
  \hline
  \hline
   \multicolumn{1}{c @{\qquad}}{systematic} &  &  \\
   \multicolumn{1}{c @{\qquad}}{error of} & $K^0$ & $\Lambda$  \\
  \hline
                                           & $\quad$+4$\quad$     & $\quad$+9$\quad$ \\
   \raisebox{1.5ex}[-1.5ex]{slopes [MeV]}  & $\quad$-7$\quad$     & $\quad$-7$\quad$ \\[0.10cm]
                                           & $\quad$+0.024$\quad$ & $\quad$+0.047$\quad$ \\
   \raisebox{1.5ex}[-1.5ex]{$dN/dy$ width} & $\quad$-0.031$\quad$ & $\quad$-0.031$\quad$ \\[0.10cm]
                                           & $\quad$+0.007$\quad$ & $\quad$+0.009$\quad$ \\
   \raisebox{1.5ex}[-1.5ex]{total yield}   & $\quad$-0.008$\quad$ & $\quad$-0.025$\quad$ \\[0.05cm]
  \hline
  \hline
 \end{tabular}
 \caption{\label{tab-3} Systematic errors on the $K^0$ and $\Lambda$ results.}
\end{table}

Furthermore, an independent analysis based on a multi-layered neural
network \cite{bre95}, also has been used to identify $\Lambda$
\cite{lop03,lop04}. Within errors the results of yields and inverse
slope parameters show a good agreement with the ones obtained from the
standard procedure.

\subsection{\label{ssec:sbal}Strangeness Balance}
In heavy-ion collisions, strange particles are produced via the strong
interaction which conserves the total amount of strangeness in the
system. At beam energies of about 2$A$~GeV the bulk part of the
produced strange particles comprises the $\Lambda$ and $\Sigma$
baryons and the kaons ($K^{\pm},K^0,\overline{K^0}$). Heavier strange
resonances or multi-strange particles have ($NN$) production
thresholds which lie significantly above this energy and thus are
strongly suppressed. Therefore, their contribution to the overall
strangeness production is negligible and will not be considered in the
following.

Due to the conservation of strangeness, the balanced sum of the
production probabilities of the above mentioned particles then has to
cancel:
\begin{equation}
\label{eqt7} 
  0 = \Sigma^++\Sigma^-+\Sigma^0+\Lambda^0+K^-+\overline{K^0}-\left(K^++K^0\right)
  \textrm{.}
\end{equation}
Here the symbols of the particles denote the respective strangeness
production probabilities.
As mentioned earlier, the $\Sigma^0$ can not be disentangled from the
$\Lambda^0$ due to its decay properties and to the detector
itself. Thus one always measures the $(\Lambda^0+\Sigma^0)$ yield. In
the case of neutral kaons, one measures the decay of $K^0_S$
into two charged pions. After acceptance and efficiency corrections
one obtains the yield of $(K^0_S+K^0_L)$ which is equal to
$(K^0+\overline{K^0})$.

\begin{table}[!hbt]
 \begin{tabular}{r @{\qquad} r @{\qquad} r @{\qquad} r}
 \hline
 \hline
   & & \multicolumn{1}{c @{\qquad}}{stat.} & \multicolumn{1}{c}{syst.} \\
  \multicolumn{1}{c @{\qquad}}{particle} & \multicolumn{1}{c @{\qquad}}{yield} & \multicolumn{1}{c @{\qquad}}{error} & \multicolumn{1}{c}{error}  \\
 \hline
  \multicolumn{1}{l}{$K^+$ \cite{bes97,men00}} & 0.075 & $\pm$~0.002 &  \\
  \multicolumn{1}{l}{$K^-$ \cite{men00}}       & 0.003 & $\pm$~0.001 &  \\[0.10cm]
  $K^0+\overline{K^0}$ & 0.078 & $\pm$~0.003 & $^{+0.007}_{-0.008}$ \\
  $\Lambda+\Sigma^0$   & 0.137 & $\pm$~0.005 & $^{+0.009}_{-0.025}$ \\[0.05cm]
 \hline
 \hline
 \end{tabular}
 \caption{\label{tab-2} Measured production yields of strange particles
   in central Ni+Ni collisions (350~mb).}
\end{table}  

Using the known yields of $K^+$ \cite{bes97,men00}, $K^-$ \cite{men00},
$K^0+\overline{K^0}$ and $\Lambda^0+\Sigma^0$ in Ni+Ni collisions at
1.93$A$~GeV beam energy given in Tab.~\ref{tab-2}, one can now determine
the missing part of the bulk strangeness, i.e. the charged strange
baryons $\Sigma^++\Sigma^-$. These baryons always decay to final
states including neutral particles and thus can not be detected with
the FOPI setup.
In order to estimate the production yield of charged $\Sigma$ baryons
we use the following relation:
\begin{equation}
\label{eqt8} 
  \Sigma^++\Sigma^- = K^++(K^0+\overline{K^0})-(\Sigma^0+\Lambda^0)-3\cdot K^-
\end{equation}
under the assumption that the number of produced $\overline{K^0}$ is
approximately equal to the number of produced $K^-$.
The result is a yield of 0.007~$\pm$~0.008~(stat.)~$^{+0.032}_{-0.017}$~(syst.)
particles per event (cf. section \ref{ssec:syse} for the determination
of the systematic errors). Thus there is very little room for the
production of charged (and due to isospin symmetry also neutral) $\Sigma$
baryons.

This result can be understood partially from the fact that due to the
mass difference between $\Lambda^0$ and $\Sigma^{\pm,0}$ of about
80~MeV, the thresholds for the direct production in $NN$ collisions
are different. While about 1.6$A$~GeV beam energy is needed to produce
a ($\Lambda^0,K^{+,0}$) pair, the creation of a
($\Sigma^{\pm,0},K^{+,0}$) pair requires energies higher than
1.8$A$~GeV. At a beam energy of 1.93$A$~GeV the $\Lambda^0$ therefore is
produced at an excess energy (in the center-of-mass system) of about
125~MeV whereas only 50~MeV are available for the $\Sigma$
baryons. Thus, one already expects a certain suppression of the
$\Sigma^{\pm,0}$ with respect to the $\Lambda^0$.

Experimental studies of $p$-$p$ collisions by the COSY-11
collaboration \cite{sew99} have shown that at (center-of-mass) excess
energies of a few MeV, the direct production of the $\Sigma^0$ is
suppressed by about a factor 30 relative to the $\Lambda^0$. In
another experiment by COSY-11, the energy dependence of the $\Sigma^+$ cross
section was investigated \cite{roz06} and was found to be enhanced by
about a factor 10 relative to the $\Lambda^0$ results. Both, the $\Sigma^0$
suppression and the $\Sigma^+$ enhancement seem to be a result of
strong final state interactions between the outgoing hyperon and the
remaining nucleon.

Evaluating the results of COSY-11 for our situation of different
excess energies for $\Lambda^0$ and $\Sigma$ production, we find a
suppression factor of about 30 for the $\Sigma^0$ with
respect to the $\Lambda^0$ and an enhancement factor of three
for the $\Sigma^+$.
While the $\Sigma^0$ suppression is fully consistent with our results
from the strangeness balance, the $\Sigma^+$ enhancement is clearly
incompatible. Transport and statistical model approaches \cite{aachn} 
calculate a $(\Sigma^++\Sigma^-)/(\Lambda^0+\Sigma^0)$ ratio of the
order of 0.35, while a value of 0.05 is found from our data.
A possible explanation could be strangeness exchange occurring in
collisions between $\Sigma^+$ and neutrons with a $\Sigma^0$ in the
final state. This channel would then increase the $\Lambda^0$ yield
and produce the observed behavior of the strangeness balance.

\subsection{\label{ssec:cdep}Centrality Dependence}
\begin{figure}[!ht]
 \includegraphics[width=\columnwidth]{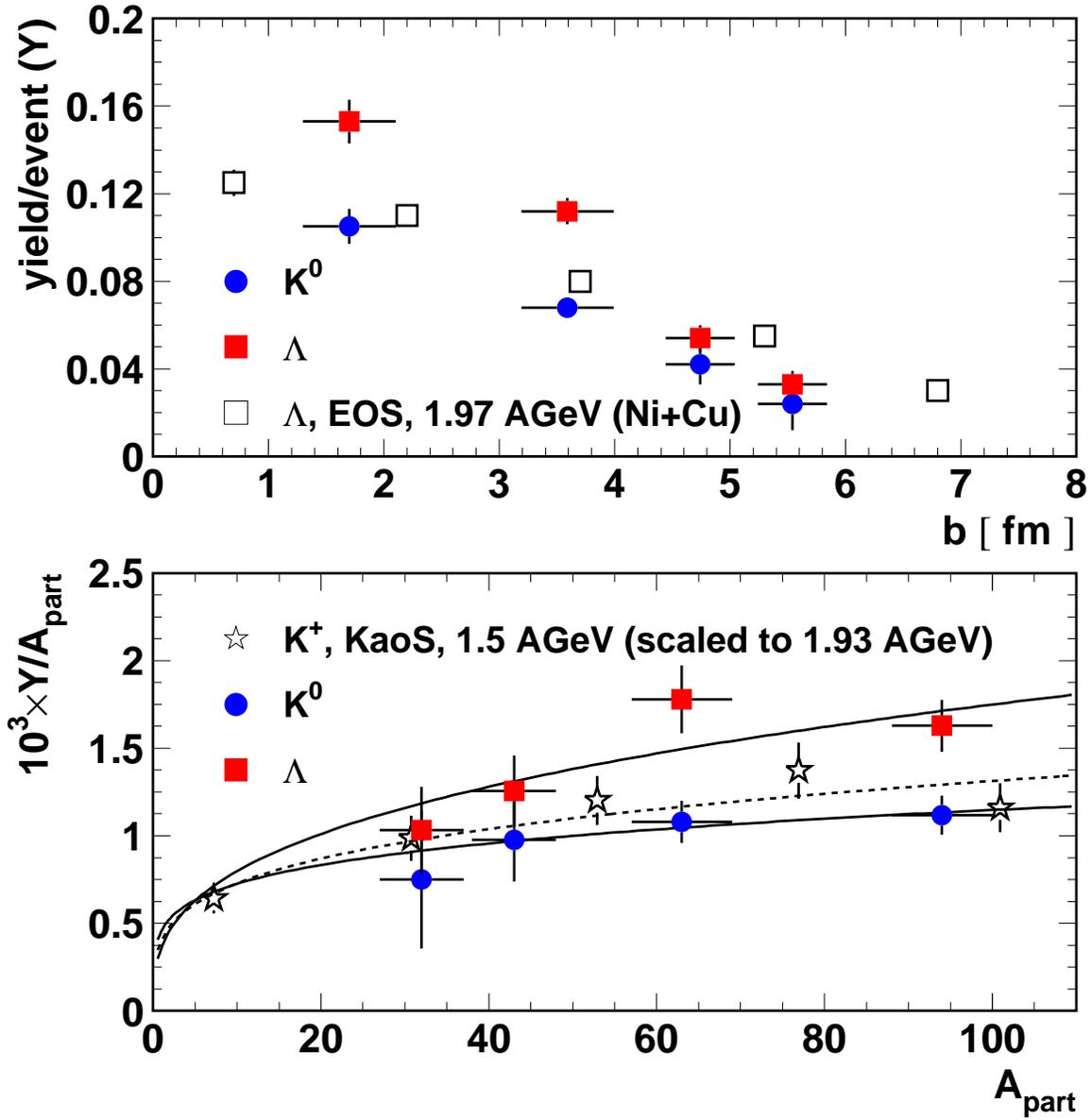}
 \caption{\label{fig:cdep} (Color online) Upper panel: Yield per event of 
 $\Lambda$ (full squares) and $K^0$ (full circles) as a function of
 the impact parameter $b$. The open squares indicate $\Lambda$ data
 from the EOS experiment \cite{jus98}.
 Lower panel: Yield per event of $\Lambda$ and $K^0$ per
 participating nucleon as function of $A_{part}$. The open stars
 denote experimental results for $K^+$ from the KaoS experiment
 \cite{for07}.}
\end{figure}

The yield per event of $K^0$ (full circles) and $\Lambda$ (full
squares) as a function of the centrality (i.e.\ impact parameter $b$,
cf.\ Tab.~\ref{tab-1}) of the reaction is presented in
Fig.~\ref{fig:cdep}.
As expected, the yield per event increases strongly with increasing 
centrality, for both particles (top panel). One observes an almost linear 
dependence of the $K^0$ and $\Lambda$ production yields on the impact 
parameter. This has been already observed at BEVALAC in Ni~+~Cu
reactions at 2$A$~GeV \cite{jus98}. Despite the qualitative agreement
of both measurements for the $\Lambda$, our data show an increase of
the production yield from the most peripheral to the most central bin
which is about 20\% higher than that from the EOS experiment.
The observed discrepancy lies within the systematic errors of the yields
of both experiments. Due to missing errors for the mean impact parameters
of the EOS centrality bins, the effects of multiplicity fluctuations and
of the poor resolution of low impact parameters on the data are unclear.
A more specific discussion of the observed discrepancy therefore is not
possible.

The total yield per event and per participating nucleon
($P/A_{part}$) is plotted as a function of $A_{part}$ (cf.\
Tab.~\ref{tab-1}) in the bottom panel of Fig.~\ref{fig:cdep}. For
both, $K^0$ and $\Lambda$, it
exhibits a similar rise with $A_{part}$, which is of the order of 50\%
from the most peripheral (MUL4) to the most central (MUL1) bin.
The dependence of $P/A_{part}$ on the number of participants can be
parameterized by
\begin{equation}
\label{eqt9} 
  \frac{P}{A_{part}} \propto \left(A_{part}\right)^{(\alpha-1)}
\end{equation}
with $\alpha_{K^0}=1.20\pm0.25$ and $\alpha_{\Lambda}=1.34\pm0.16$ for
$K^0$ and $\Lambda$, respectively.

Within errors the $K^0$ data agree with results on the centrality
dependence of the $K^+$ production yields measured by the KaoS
collaboration for the same system \cite{for07} (star symbols). Note
that for comparison, the KaoS data measured at 1.5$A$~GeV have
been scaled by a factor of 3.80 since it was shown that the
increase of the $K^+$ production yield with increasing beam energy
($E_{beam}$) is proportional to $E_{beam}^{5.3\pm0.2}$
\cite{bar97}. Similar trends were also evidenced for $K^+$ and $K^-$
measured at lower beam energies in Au~+~Au and Ni~+~Ni collisions
\cite{for03,for07}.

The variation of the production yield per $A_{part}$ as a function of 
$A_{part}$ can be interpreted as a signature of multiple collisions 
which are dominantly contributing to the production of strange
particles near threshold \cite{bar97,mic94}. Moreover, the similar
trends evidenced for $K^0$ ($K^+$) and $\Lambda$ allow to confirm that
these particles are produced according to the same creation
mechanisms: $B~B~\rightarrow~B~Y~K$ and $\pi~B~\rightarrow~Y~K$, where
$B$ stands for a nucleon or a $\Delta$ ($N^*$) resonance and $Y$ for a
$\Lambda$ or a $\Sigma$ hyperon, in the 1-2$A$~GeV energy range
\cite{fuc97}.

Systematically higher values of $\alpha$ are expected when excluding
the most central data point from the fit using Eq.~\ref{eqt9}. For the
$K^0$ one finds $\alpha'_{K^0}$~=~1.38$\pm$0.54 and
$\alpha'_{\Lambda}$~=~1.84$\pm$0.35 for the $\Lambda$, respectively.
The difference between these results and the ones from the full fit
mentioned above could imply a saturation of the strangeness production
in very central collisions. The effects of the multi-step processes
could be partially canceled by the reshuffling of strange particles
by absorption in the large volume of the fireball.

\section{\label{sec:concl}Conclusions}
We have reported on new results from the FOPI experiment concerning
the production of neutral strange particles, $K^0$ and $\Lambda$, in
Ni~+~Ni collisions at 1.93$A$~GeV. A large statistics of high quality
$K^0_S$ and $\Lambda$ candidates, with a signal-to-background ratio of
the order of one to two, have been reconstructed. Phase-space
distributions have been measured down to low transverse momenta and in
the full rapidity range. Differential yield distributions and inverse
slope parameter distributions have been found similar for $K^0$ and
$K^+$, confirming that their production mechanisms are identical and
that there is no measurable effect of the isospin difference between
the two particles.
The rapidity distributions of protons and $\Lambda$ exhibit different
trends due to an incomplete stopping in the studied system. Regarding
the overall strangeness production, the results request a strong
suppression of the charged $\Sigma$ hyperon production which is not
seen in transport or thermal model calculations. The mechanism for
this has yet to be found.
For the first time at SIS energies, the total production yield of $K^0$
and $\Lambda$ per participant is studied as a function of the
centrality of the reaction. The increase of the yield with rising
number of participants, similar for both $K^0$ and $\Lambda$,
indicates that multiple collisions contribute to the production of
neutral strange particles near threshold.
  
All observed trends confirm that $K^0$($K^+$) and $\Lambda$ production 
mechanisms are strongly correlated.
Detailed comparisons to transport models have to be done in order to
check the consistency of the data with the hypothesis of modified
in-medium properties of strange hadrons.

\subsection*{Acknowledgments}
This work was partly supported by the German Federal Ministry of Education
and Research (BMBF) under grant No. 06HD154, by the Gesellschaft f\"ur
Schwerionenforschung (GSI) under grant No. HD-HER, by the Korea Research
Foundation grant (KRF-2005-041-C00120), by the European Comission
under the 6th Framework Programme under the 'Integrated Infrastructure
Initiative on: Strongly Interacting Matter (Hadron Physics)' (Contract
no. RII3-CT-2004-506078) and by the agreement between GSI and IN2P3/CEA.

\end{document}